\documentclass{aastex631}

\accepted{July 11, 2022}

\shorttitle{ALMA Pleiades}
\shortauthors{Sullivan et al.}

\begin{document}

\title{An ALMA 1.3 millimeter Search for Debris Disks around Solar-type Stars in the Pleiades}

\correspondingauthor{David Wilner}
\email{dwilner@cfa.harvard.edu}

\author{Devin Sullivan}
\affiliation{Center for Astrophysics \textbar\ Harvard \& Smithsonian,
60 Garden St., Cambridge, MA 02138, USA}

\author[0000-0003-1526-7587]{David Wilner}
\affiliation{Center for Astrophysics \textbar\ Harvard \& Smithsonian,
60 Garden St., Cambridge, MA 02138, USA}

\author[0000-0003-4705-3188]{Luca Matr\`a}
\affiliation{School of Physics, Trinity College Dublin, The University of Dublin, College Green, Dublin 2, Ireland}
\affiliation{Centre for Astronomy, School of Physics, 
National University of Ireland Galway, University Road, Galway, Ireland H91 TK33} 

\author[0000-0001-9064-5598]{Mark C. Wyatt}
\affiliation{Institute of Astronomy, 
University of Cambridge, Madingley Road, Cambridge CB3 0HA, UK}

\author[0000-0003-2253-2270]{Sean M. Andrews}
\affiliation{Center for Astrophysics \textbar\ Harvard \& Smithsonian, 
60 Garden St., Cambridge, MA 02138, USA}

\author[0000-0001-7891-8143]{Meredith A. MacGregor}
\affiliation{Department of Astrophysical and Planetary Sciences, University of Colorado, 
2000 Colorado Avenue, Boulder, CO 80309, USA}

\author[0000-0003-3017-9577]{Brenda Matthews}
\affiliation{Herzberg Astronomy \& Astrophysics Research Centre, National Research Council of Canada, 
5071 West Saanich Road, Victoria, BC V9E 2E7, Canada}

\begin{abstract}
Millimeter emission from debris disks around stars of different ages provides constraints on
the collisional evolution of planetesimals. 
We present ALMA 1.3 millimeter observations of a sample of 76 Solar-type stars in the 
$\sim 115$~Myr-old Pleiades star cluster. These ALMA observations complement previous  
infrared observations of this 
sample  by  providing sensitivity to emission from circumstellar dust at lower temperatures, corresponding 
to debris 
at radii comparable to the Kuiper Belt and beyond. The observations obtain a beam size
of $1\farcs5$ (200 au) and a median rms noise of $54$~$\mu$Jy~beam$^{-1}$, which  corresponds to a fractional
luminosity  $L_{dust}/L_{star} \sim 10^{-4}$ for 40~K dust for a typical star in the sample. The ALMA images 
show no significant detections of the targeted stars. 
We interpret these limits in the context of a steady-state collisional cascade model for debris disk 
evolution that provides a good description of observations of the field population near the Sun
but is not well-calibrated on younger populations.
The ALMA non-detections of the Pleiades systems are compatible with the disk flux predictions of this model. 
We find no high fractional luminosity outliers from these ALMA data that could be associated with 
enhanced collisions resulting from activity not accounted for by steady-state evolution.
However, we note that two systems (HII 1132 and HD 22680)
show 24~$\mu$m excess much higher than the predictions of this model,
perhaps due to unusually high dust production from dynamical events involving planets. 
\end{abstract}

\section{Introduction} 
\label{sec:newintro}
The collisions between planetesimals orbiting main-sequence stars produce dust heated by stellar radiation 
that is manifested as debris disk emission. 
Debris disk detection statistics together with trends of dust excess with stellar age probe the evolution of planetesimal belts, 
which may be influenced by planets (see reviews by \citet{Wyatt2008,Matthews+2014,Hughes+2018}).
Studies of 24 and 70 $\mu$m dust emission around stars with a range of ages and spectral types show that debris disk detection frequency 
and fractional luminosity ($L_{dust}/L_{star}$) rise over perhaps 10’s of Myr and then decline slowly over 
100’s of Myr \citep{Rieke+2005,Su+2006,Carpenter+2009}.
These data can be explained in a theoretical framework that invokes an early phase of runaway growth and 
merging of planetesimals  \citep{KenyonBromley2008} followed by a steady state destructive collisional cascade \citep{DominikDecin2003}.
In this picture, planetesimals collide and fragment into smaller objects that in turn collide and fragment,
ending in circumstellar dust small enough to be removed by stellar winds and radiation.
Because collisional depletion is faster for smaller disks and more massive disks, the wide dispersion of 
fractional luminosity and  dust temperature observed at a given age may be accounted for by variations in the initial planetesimal belt masses and radii  \citep{Wyatt+2007a,Lohne+2008}.
This standard collisional evolution model for debris disks predicts a maximum fractional
luminosity of $10^{-3}$ at early times, in line with the most extreme systems known \citep{Wyatt+2007b}.

Observational constraints on evolutionary models are conspicuously sparse for debris disks around Solar-type stars 
at large radii, beyond 10 au, for ages near 100 Myr, i.e. 
older than nearby moving groups but younger than the nearby field. 
This is a critical epoch when young planetary systems may be dynamically active. 
Such systems are especially relevant to provide context for the evolution of our Kuiper Belt, 
which apparently was destabilized by the outward migration of Neptune and underwent a dramatic dynamical depletion, 
from an initial mass  of more than 10~M$_{\Earth}$ at radii 30 to 50 au, to less than 0.1 M$_{\Earth}$ today; 
modeling suggests that the corresponding fractional luminosity decreased from $10^{-3}$ to $10^{-7}$ \citep{Booth+2009}.
This is also the epoch of the formation of the Earth's moon \citep[e.g. see review by][]{Asphaug14}.
Sensitive surveys of the (older) field population of Solar-type stars (FGK type) within 20 pc show the occasional
system (5 of 275) with debris at large radii and high fractional luminosity ($>5\times10^{-5}$) 
whose presence is not predicted by standard steady-state collisional evolution models \citep{Sibthorpe+2018}.
A plausible hypothesis for such large excess is an ongoing system-wide dynamical event involving 
planets, akin to the one that shaped the Kuiper Belt.
Another possibility is that these systems follow the ``late bloomer'' path
revealed by coagulation models, where protoplanetary disk rings with nearly completely 
efficient planetesimal formation slowly rise to these high fractional luminosities over 
$\sim1$~Gyr \citep{Najita+2022}.

The Pleiades cluster has long provided a benchmark for debris disk evolution studies because of its 
combination of (1) age: $115 \pm 10$ Myr \citep{Stauffer+1998},
the epoch of maximum dust production rate in ``self-stirred'' evolutionary models  \citep{KenyonBromley2008},
(2) proximity: $133.5\pm1.2$ pc \citep{Melis+2014,GAIA2018}, for maximum sensitivity, and 
(3) a sufficiently large sample: $\mathcal{O}(100)$ Solar-type stars,  to obtain meaningful statistics to compare 
with evolutionary models. No other star cluster satisfies these criteria.
Sensitive searches for warm ($>100$ K) circumstellar dust in the Pleiades were performed in a series of Spitzer 24 $\mu$m surveys 
that include members of spectral type F5 to K1  \citep{Stauffer+2005,Gorlova+2006,Sierchio+2010}.
For systems observed at 24 $\mu$m, 23/71 (32\%) show excesses greater than 10\% above the photospheres, 
attributed to dusty debris at blackbody radii from $\sim1$ to 10~au. 
By comparison, only 3/164 (2\%) of the Solar-type stars in the 750 Myr old Praesepe cluster 
\citep[mean distance 187.3~pc,][]{Lodieu+2019}
show 24 $\mu$m excesses, 
as dust production declines with time due to collisional depletion in the inner regions \citep[e.g.][]{Gaspar+2009}. 
All models that explain debris disk evolution predict that the Pleiades should harbor a population of 
colder ($< 100$ K) debris belts at larger radii ($> 10$ au).

There are few existing constraints on cold dust belts in the Pleiades. 
Emission at 24 $\mu$m is on the Wien side of the spectral peak for  dust at radii of 10’s of au 
and therefore exponentially depressed. 
A subset of 20 Pleiades systems was observed by {\em Spitzer} at 70 $\mu$m as a part of the 
{\em FEPS} program, which targeted 
328 Solar-type stars spanning ages of 3 Myr to 3 Gyr to study debris disk evolution \citep{Carpenter+2009}.
No excess emission was detected from these 20 Pleiades targets at 70 $\mu$m,  
but with only marginal sensitivity to the outer belts of interest \citep{Stauffer+2005}.
Millimeter wavelengths provide 
direct access to cold dust, but millimeter observations 
of  the Pleiades have yet  to reach relevant levels  of sensitivity to $L_{dust}/L_{star}$ 
\citep{ZuckermanBecklin1993,Greaves+2009,Roccatagliata+2009}.

The Survey of Nearby Stars (SONS) with the JCMT provided 0.85 (and 0.45)~millimeter follow-up of 
debris disks near the Sun known from far-infrared surveys, resulting in 31 detections 
of FGK stars \citep{Holland+2017}. Nearly all systems with ages $<50$~Myr were detected, with lower detection 
rates at older ages, in line with expectations for collisional evolution. 
Only 6 systems with ages in the range 50 to 200~Myr were detected. 
Now, the unprecendented sensitivity of ALMA allows for unbiased surveys of millimeter emission
from circumstellar to constrain debris disk evolutionary models. 
For example, \citet{Lovell+2021} used ALMA to detect 4/30 Class III systems in the Lupus clouds and
concluded that the debris belts found around older stars are already present at an age of $\sim2$~Myr. 

In this paper, we report the results of an ALMA survey of Solar-type stars in the Pleiades cluster 
at 1.3 millimeters wavelength. 
These observations improve on previous millimeter observations by about an order of magnitude in 
sensitivity and for a much larger sample, and they complement previous infrared observations of the 
same stars. 
In \S\ref{sec:observations}, we describe the details of the ALMA observations.
In \S\ref{sec:results}, we present the results for the sample.
In \S\ref{sec:discussion}, we discuss these results in the framework of steady-state collisional 
evolution developed by \citet{WyattDent2002} and previously applied to Solar-type stars by 
\citet{Kains+2011} and \citet{Sibthorpe+2018}.
Finally, \S\ref{sec:conclusions} summarizes the conclusions from this study.

\section{Observations}
\label{sec:observations}
\subsection{ALMA Sample Selection}
We based the ALMA survey targets on the sample of 76 Solar-type Pleiades stars with reliable 
24 $\mu$m photometry compiled by \citet{Sierchio+2010}.\footnote{We note two modifications of 
the ALMA sample from the compilation of \citet{Sierchio+2010}:
AK~1B~146 (HD~23170, HII~177) was not included as it is not a Pleiades member \citep{Torres+2021},
and HII~1776 was included as it was in the FEPS sample observed at 70~$\mu$m by \citet{Stauffer+2005}.} 
The sample was derived from the work of  \citet{Stauffer+2005} that determined each star to be a  likely Pleiades cluster 
member based on radial velocities, proper motions, chromospheric and coronal activity indicators, and lithium abundance. 
These stars have colors $1.5 < V - K_S < 2.15$, corresponding to spectral types F5 to K1.  
None of the stars in this sample shows evidence for hot ($> 300$ K) dust, from stringent limits of 
excess emission at  wavelengths $\lesssim8$~$\mu$m. This is consistent with the expected clearing of 
inner disk regions ($< 1$ au) by collisions over $\sim100$ Myr (and perhaps also by planets).
Table~\ref{tab:targets} lists the target stars and their coordinates, 
together with the stellar effective temperatures and luminosities from
the analysis provided in the {\em Gaia} DR2 catalog \citep{GAIA2018} where available
Table~\ref{tab:targets} also lists estimates of the stellar mass based on the luminosity-mass relation 
of \citet{DemircanKahraman1991} for zero-age main sequence stars in this mass range. 
The last column lists the 24~$\mu$m excess ratio ($F_{obs}/F_{star}$) determined by 
\citet{Sierchio+2010}, with significant detections of circumstellar dust emission indicated in bold
(excess ratio $>1.1$, i.e. exceeding $3\sigma$).
The 20 systems observed at 70~$\mu$m by \citet{Stauffer+2005} are at the end of Table~\ref{tab:targets}
(demarcated by a line).

\startlongtable
\begin{deluxetable}{llccccrcc}
\tablecaption{ALMA Pleiades Targets
\label{tab:targets}
}
\tablewidth{0pt}
\tabletypesize{\scriptsize}
\tablehead{
\colhead{Name} & \colhead{Alternate} & \colhead{RA} & \colhead{Dec} & \colhead{Distance\tablenotemark{a}} &  \colhead{$T_{eff}$\tablenotemark{b}} & \colhead{$L_{star}$\tablenotemark{b}} & \colhead{$M_{\star}$} 
& \colhead{24~$\mu$m}\\
\colhead{} & \colhead{Name} & \colhead{(J2000)} & \colhead{(J2000)} & \colhead{(pc)} & \colhead{(K)} & \colhead{($L_{\odot}$)} & \colhead{($M_{\odot}$)} 
& \colhead{excess ratio\tablenotemark{c}}
} 
\startdata
HIP 17317 & AK IA 36 & 3:42:24.05 & 22:25:15.2 & $137.4_{-1.0}^{+1.0}$ & $5626_{-113}^{+101}$ & $1.19_{-0.01}^{+0.01}$ & 1.04 & 1.03\\
BD+21 508 & AK IA 56 & 3:43:31.15 & 22:09:29.2 & $135.0_{-1.1}^{+1.2}$ & $5468_{-123}^{+379}$ & $1.14_{-0.01}^{+0.01}$ & 1.03 & {\bf 1.13} \\
HD 23312 & AK IA 76 & 3:44:58.94 & 22:01:56.0 & $141.1_{-1.2}^{+1.2}$ & $6253_{-147}^{+134}$ & $2.66_{-0.03}^{+0.03}$ & 1.32 & {\bf 1.26} \\
HD 24463 & AK IA 317 & 3:54:21.62 & 24:04:31.6 & $135.3_{-1.0}^{+1.0}$ & $6136_{-190}^{+201}$ & $1.97_{-0.02}^{+0.02}$ & 1.21 & 1.01\\
HD 22627 & AK IB 7 & 3:39:09.15 & 24:22:03.3 & $143.2_{-1.0}^{+1.0}$ & $5924_{-92}^{+81}$ & $2.01_{-0.02}^{+0.02}$ & 1.22 & {\bf 1.12}\\
HIP 17044 & AK IB 8 & 3:39:13.51 & 24:27:58.6 & $141.7_{-0.9}^{+0.9}$ & $5733_{-761}^{+181}$ & $1.24_{-0.01}^{+0.01}$ & 1.06 & 1.04\\
HD 23598 & AK IB 365 & 3:47:20.91 & 25:31:32.1 & $136.2_{-0.9}^{+1.0}$ & $5892_{-266}^{+113}$ & $1.84_{-0.02}^{+0.02}$ & 1.18 & 1.01\\
HD 24086 & AK IB 590 & 3:51:06.36 & 25:35:40.0 & $139.0_{-1.3}^{+1.3}$ & $6430_{-149}^{+270}$ & $3.55_{-0.04}^{+0.04}$ & 1.44 & 0.99\\
HD 23975 & AK IB 560 & 3:50:17.70 & 25:22:45.5 & $137.3_{-1.2}^{+1.3}$ & $6155_{-212}^{+283}$ & $2.17_{-0.03}^{+0.03}$ & 1.24 & {\bf 1.34} \\
HD 22444 & AK II 34 & 3:37:24.07 & 22:21:02.6 & $132.3_{-1.5}^{+1.6}$ & $6135_{-101}^{+102}$ & $3.04_{-0.04}^{+0.04}$ & 1.37 & 1.06\\
TYC 1798-465-1 & AK II 359 & 3:37:34.93 & 24:14:10.8 & $155.4_{-1.2}^{+1.2}$ & $5713_{-252}^{+156}$ & $1.21_{-0.01}^{+0.01}$ & 1.05 & 0.99\\
HIP 16979 & AK II 383 & 3:38:22.61 & 22:29:58.0 & $137.4_{-0.9}^{+0.9}$ & $5880_{-129}^{+118}$ & $1.50_{-0.01}^{+0.01}$ & 1.12 & {\bf 1.12} \\
HD 22680 & AK II 437 & 3:39:41.22 & 23:17:26.3 & $140.5_{-1.1}^{+1.1}$ & $6057_{-217}^{+181}$ & $1.75_{-0.02}^{+0.02}$ & 1.17 & {\bf 3.79} \\
HIP 16639 & AK III 288 & 3:34:07.34 & 24:20:39.1 & $137.2_{-0.9}^{+0.9}$ & $6086_{-171}^{+160}$ & $2.36_{-0.02}^{+0.02}$ & 1.27 & 0.99\\
HII 25 & HD 23061 & 3:42:55.15 & 24:29:34.2 & $138.4_{-1.0}^{+1.0}$ & $6197_{-137}^{+464}$ & $2.54_{-0.02}^{+0.02}$ & 1.30 & {\bf 1.12} \\
HII 102 & TYC 1799-118-1 & 3:43:24.56 & 23:13:32.5 & $138.8_{-0.7}^{+0.7}$ & $5338_{-60}^{+529}$ & $1.09_{-0.01}^{+0.01}$ & 1.02 & 1.04\\
HII 1132 & HD 23514 & 3:46:38.43 & 22:55:10.4 & $138.2_{-1.0}^{+1.0}$ & $6269_{-196}^{+392}$ & $2.68_{-0.03}^{+0.03}$ & 1.32 & {\bf 17.44} \\
HII 1139 & HD 23513 & 3:46:40.02 & 23:06:36.3 & $137.2_{-1.1}^{+1.1}$ & $6365_{-150}^{+179}$ & $2.67_{-0.03}^{+0.03}$ & 1.32 & 1.09\\
HII 1766 & HD 23732 & 3:48:16.91 & 25:12:53.6 & $137.8_{-1.4}^{+1.5}$ & $6614_{-336}^{+430}$ & $3.19_{-0.04}^{+0.04}$ & 1.39 & {\bf 1.60} \\
HII 2172 & HD 282965 & 3:49:11.77 & 24:38:10.9 & $139.1_{-1.1}^{+1.1}$ & $5809_{-190}^{+50}$ & $1.10_{-0.01}^{+0.01}$ & 1.02 & {\bf 1.10} \\
HII 3031 & HD 24132 & 3:51:27.25 & 24:31:06.2 & $135.9_{-1.2}^{+1.2}$ & $6753_{-201}^{+239}$ & $4.25_{-0.05}^{+0.05}$ & 1.51 & 1.01 \\
TYC 1802-95-1 & PELS 7 & 3:34:47.23 & 26:05:40.3 & $177.2_{-1.7}^{+1.8}$ & $5596_{-105}^{+163}$ & $1.79_{-0.02}^{+0.02}$ & 1.18 & 0.99 \\
HIP 17020 & PELS 20 & 3:38:56.90 & 24:34:10.4 & $139.1_{-1.3}^{+1.3}$ & $5723_{-170}^{+127}$ & $1.02_{-0.01}^{+0.01}$ & 1.00 & {\bf 1.19} \\
HIP 17245 & PELS 23 & 3:41:36.19 & 25:37:08.6 & $138.5_{-1.0}^{+1.0}$ & $5800_{-175}^{+124}$ & $1.54_{-0.01}^{+0.01}$ & 1.12 & 1.06 \\
HIP 17125 & PELS 25 & 3:40:03.11 & 27:44:24.9 & $138.8_{-2.5}^{+2.6}$ & $6220_{-133}^{+95}$ & $2.26_{-0.05}^{+0.05}$ & 1.26 & 1.02\\
BD+21 516 & PELS 40 & 3:45:09.91 & 21:42:15.5 & $127.3_{-0.8}^{+0.8}$ & $5837_{-156}^{+267}$ & $1.41_{-0.01}^{+0.01}$ & 1.10 & 1.03 \\
HIP 18544 & PELS 86 & 3:58:01.72 & 20:40:35.5 & $125.1_{-1.1}^{+1.1}$ & $6185_{-102}^{+127}$ & $2.24_{-0.03}^{+0.03}$ & 1.26 & 1.10 \\
BD+23 455 & PELS 121 & 3:27:42.07 & 23:48:12.4 & $195.1_{-1.7}^{+1.7}$ & $5676_{-111}^{+155}$ & $2.58_{-0.03}^{+0.03}$ & 1.31 & 0.97 \\
HIP 16753 & PELS 124 & 3:35:31.73 & 22:49:24.0 & $131.5_{-0.8}^{+0.8}$ & $6094_{-135}^{+196}$ & $1.64_{-0.01}^{+0.01}$ & 1.15 & 1.06 \\
BD+26 592 & PELS 128 & 3:39:53.74 & 26:43:00.3 & $221.7_{-2.6}^{+2.7}$ & $5633_{-101}^{+69}$ & $3.40_{-0.06}^{+0.06}$ & 1.42 & 1.05 \\
TYC 1256-516-1 & PELS 135 & 3:45:01.68 & 19:33:32.8 & $137.4_{-1.0}^{+1.0}$ & $6190_{-167}^{+257}$ & $2.67_{-0.03}^{+0.03}$ & 1.32 & 1.10 \\
HIP 18091 & PELS 146 & 3:52:00.83 & 19:35:47.8 & $143.1_{-0.8}^{+0.8}$ & $5593_{-205}^{+234}$ & $1.16_{-0.01}^{+0.01}$ & 1.04 & {\bf 1.43} \\
HD 23935 & PELS 150 & 3:49:52.94 & 25:38:49.9 & $138.5_{-0.8}^{+0.8}$ & $6008_{-151}^{+163}$ & $2.41_{-0.02}^{+0.02}$ & 1.28 & {\bf 1.26} \\
BD+22 617C & PELS 173 & 4:00:53.14 & 23:11:37.5 & $184.3_{-2.2}^{+2.2}$ & $6457_{-112}^{+205}$ & $3.95_{-0.06}^{+0.06}$ & 1.48 & 1.18 \\
HIP 18955 & PELS 174 & 4:03:44.20 & 22:56:38.5 & $137.0_{-1.1}^{+1.1}$ & $5767_{-266}^{+144}$ & $2.20_{-0.02}^{+0.02}$ & 1.25 & 1.06\\
HIP 17316 & TrS 42 & 3:42:24.03 & 21:28:23.6 & $129.1_{-0.8}^{+0.8}$ & $5954_{-148}^{+274}$ & $1.62_{-0.01}^{+0.01}$ & 1.14 & {\bf 1.10} \\
HD 24302 & Tr 60 & 3:52:53.50 & 24:42:55.7 & $140.5_{-1.2}^{+1.2}$ & $6376_{-108}^{+181}$ & $2.68_{-0.03}^{+0.03}$ & 1.32 & 1.01 \\
HII 293 & V* V1169 Tau & 3:44:13.95 & 24:46:44.9 & $136.1_{-1.0}^{+1.0}$ & $5503_{-98}^{+145}$ & $0.82_{-0.01}^{+0.01}$ & 0.94 & 0.98\\
HII 405 & HD 23269 & 3:44:40.78 & 24:49:05.8 & $134.5_{-1.0}^{+1.0}$ & $6072_{-86}^{+68}$ & $1.74_{-0.02}^{+0.02}$ & 1.17 & 0.98 \\
HII 489 & TYC 1803-808-1 & 3:44:56.40 & 24:25:57.4 & $134.2_{-1.2}^{+1.2}$ & $5693_{-189}^{+464}$ & $1.06_{-0.01}^{+0.01}$ & 1.01 & {\bf 1.20} \\
HII 571 & TYC 1803-1156-1 & 3:45:15.39 & 25:17:21.2 & $134.4_{-0.9}^{+0.9}$ & $5167_{-127}^{+230}$ & $0.57_{-0.01}^{+0.01}$ & 0.84 & 1.05\\
HII 727 & V* V855 Tau & 3:45:40.20 & 24:37:37.3 & $131.6_{-1.1}^{+1.1}$ & $5927_{-60}^{+156}$ & $1.89_{-0.02}^{+0.02}$ & 1.19 & 1.00 \\
HII 739 & V* V969 Tau & 3:45:42.15 & 24:54:20.7 & $184.8_{-20.9}^{+27.0}$ & $5820_{-170}^{+480}$ & $4.40_{-0.58}^{+0.58}$ & 1.53 & 1.00 \\
HII 923 & BD+22 548 & 3:46:10.07 & 23:20:23.0 & $122.5_{-0.9}^{+0.9}$ & $5775_{-108}^{+67}$ & $1.13_{-0.01}^{+0.01}$ & 1.03 & 1.06 \\
HII 996 & V* V1045 Tau & 3:46:22.69 & 24:34:11.6 & $138.7_{-0.9}^{+0.9}$ & $5775_{-223}^{+317}$ & $1.10_{-0.01}^{+0.01}$ & 1.02 & {\bf 1.21} \\
HII 1117 & HD 282975 & 3:46:37.73 & 23:47:15.0 & $134.6_{-0.9}^{+0.9}$ & $5535_{-116}^{+45}$ & $1.34_{-0.01}^{+0.01}$ & 1.08 & 1.04 \\
HII 1309 & HD 23584 & 3:47:10.09 & 24:16:35.1 & $136.1_{-1.1}^{+1.2}$ & $6165_{-71}^{+175}$ & $2.45_{-0.03}^{+0.03}$ & 1.29 & 0.97 \\
HII 1338 & HD 23608 & 3:47:16.59 & 24:07:41.5 & $95.1_{-8.4}^{+10.2}$ & $6355_{-105}^{+141}$ & $2.34_{-0.24}^{+0.24}$ & 1.27 & 0.96 \\
HII 1514 & HD 282967 & 3:47:40.45 & 24:21:52.5 & $136.2_{-1.2}^{+1.2}$ & $5629_{-136}^{+75}$ & $0.97_{-0.01}^{+0.01}$ & 0.98 & 1.07 \\
HII 1613 & HD 282973 & 3:47:52.51 & 23:56:28.4 & $134.3_{-1.0}^{+1.0}$ & $6003_{-95}^{+121}$ & $1.68_{-0.02}^{+0.02}$ & 1.15 & 1.01 \\
HII 1726 & HD 23713 & 3:48:07.12 & 24:08:30.9 & \nodata & $6444_{-446}^{+455}$ & \nodata & \nodata & 1.04 \\
HII 1797 & BD+23 551 & 3:48:16.91 & 23:38:12.5 & $140.4_{-0.9}^{+0.9}$ & $5886_{-64}^{+84}$ & $1.49_{-0.01}^{+0.01}$ & 1.11 & {\bf 1.52} \\
HII 1856 & HD 282971 & 3:48:26.20 & 24:02:53.5 & $135.0_{-1.1}^{+1.1}$ & $5956_{-103}^{+93}$ & $1.49_{-0.02}^{+0.02}$ & 1.11 & 0.94\\
HII 1912 & HD 23778 & 3:48:34.81 & 24:10:51.2 & $221.4_{-41.8}^{+66.8}$ & $6256_{-173}^{+365}$ & $8.59_{-1.76}^{+1.76}$ & 1.86 & 0.97\\
HII 1924 & Cl Melotte 22 1924 & 3:48:34.54 & 23:26:04.4 & $136.8_{-0.9}^{+0.9}$ & $5741_{-93}^{+177}$ & $1.16_{-0.01}^{+0.01}$ & 1.04 & 1.04 \\
HII 2027 & Cl Melotte 22 2027 & 3:48:48.95 & 24:16:02.8 & $143.0_{-1.0}^{+1.0}$ & $5066_{-82}^{+363}$ & $0.88_{-0.01}^{+0.01}$ & 0.96 & 0.97\\
\hline
HII 120 & TYC 1799-102-1 & 3:43:31.98 & 23:40:25.7 & $135.7_{-1.0}^{+1.1}$ & $5524_{-94}^{+688}$ & $0.78_{-0.01}^{+0.01}$ & 0.92 & 1.04 \\
HII 152 & V* V963 Tau & 3:43:37.77 & 23:32:08.7 & $136.0_{-1.0}^{+1.0}$ & $5502_{-229}^{+161}$ & $0.83_{-0.01}^{+0.01}$ & 0.94 & {\bf 1.12} \\
HII 173 & Cl Melotte 22 173 & 3:43:48.44 & 25:11:23.6 & $130.4_{-1.6}^{+1.6}$ & $5206_{-159}^{+305}$ & $0.76_{-0.01}^{+0.01}$ & 0.91 &  1.00 \\
HII 174  & V* V1271 Tau & 3:43:48.38 & 25:00:14.8 & $136.3_{-1.0}^{+1.0}$ & $4991_{-32}^{+60}$ & $0.41_{-0.01}^{+0.01}$ & 0.76 & 0.99 \\ 
HII 250 & Cl Melotte 22 250 & 3:44:04.26 & 24:59:22.5 & $129.8_{-1.1}^{+1.1}$ & $5518_{-135}^{+166}$ & $0.76_{-0.01}^{+0.01}$ & 0.91 & {\bf 1.12} \\
HII 314 & V* V1038 Tau & 3:44:20.19 & 24:47:45.4 & $134.2_{-0.9}^{+0.9}$ & $5431_{-107}^{+87}$ & $0.90_{-0.01}^{+0.01}$ & 0.96 & 1.02 \\
HII 514 & Cl Melotte 22 514 & 3:45:04.03 & 25:15:27.4 & $135.3_{-0.8}^{+0.9}$ & $5516_{-71}^{+398}$ & $0.85_{-0.01}^{+0.01}$ & 0.95 & {\bf 1.19} \\
HII 1015 & HD 282952 & 3:46:27.37 & 25:08:07.1 & $135.0_{-1.0}^{+1.0}$ & $5626_{-117}^{+101}$ & $0.95_{-0.01}^{+0.01}$ & 0.98 & 0.98 \\
HII 1101 & HD 282954 & 3:46:38.80 & 24:57:33.8 & $136.3_{-1.0}^{+1.0}$ & $5900_{-271}^{+206}$ & $1.26_{-0.01}^{+0.01}$ & 1.06 &  {\bf 1.48} \\
HII 1182 & BD+22 552 & 3:46:47.08 & 22:54:51.5 & $138.5_{-1.0}^{+1.0}$ & $5590_{-225}^{+264}$ & $1.06_{-0.01}^{+0.01}$ & 1.01 &  1.03 \\
HII 1200 & BD+22 553 & 3:46:50.56 & 23:14:20.3 & $153.2_{-1.5}^{+1.6}$ & $5946_{-79}^{+94}$ & $2.14_{-0.03}^{+0.03}$ & 1.24 &  {\bf 1.15} \\
HII 1776 & HD 282958 & 3:48:17.72 & 25:02:51.4 & $133.5_{-1.0}^{+1.0}$ & $5499_{-107}^{+101}$ & $0.69_{-0.01}^{+0.01}$ & 0.89 & 1.02 \\
HII 2147 & V* V1282 Tau & 3:49:06.14 & 23:46:51.8 & $138.2_{-1.0}^{+1.0}$ & $5305_{-199}^{+738}$ & $1.02_{-0.01}^{+0.01}$ & 1.00 &  1.00 \\
HII 2278 & HD 282960 & 3:49:25.75 & 24:56:14.5 & \nodata & $5080_{-81}^{+324}$ & \nodata & \nodata &  0.98 \\
HII 2506 & BD+22 574 & 3:49:56.52 & 23:13:06.2 & $146.5_{-1.1}^{+1.1}$ & $5849_{-92}^{+104}$ & $1.42_{-0.01}^{+0.01}$ & 1.10 & 1.00 \\
HII 2644 & Cl Melotte 22 2644 & 3:50:20.90 & 24:28:00.3 & $137.4_{-1.0}^{+1.1}$ & $5469_{-62}^{+166}$ & $0.63_{-0.01}^{+0.01}$ & 0.87 & 1.01 \\
HII 2786 & V* V1175 Tau & 3:50:40.11 & 23:55:58.1 & $139.1_{-1.2}^{+1.2}$ & $5852_{-103}^{+199}$ & $1.25_{-0.01}^{+0.01}$ & 1.06 &  0.99 \\
HII 2881 & V* V1176 Tau & 3:50:54.32 & 23:50:05.6 & $147.1_{-13.5}^{+16.5}$ & $4917_{-57}^{+122}$ & $0.63_{-0.07}^{+0.07}$ & 0.87 &  1.00 \\
HII 3097 & Cl Melotte 22 3097 & 3:51:40.48 & 24:58:58.6 & $134.6_{-1.3}^{+1.3}$ & $5441_{-143}^{+184}$ & $0.74_{-0.01}^{+0.01}$ & 0.91 & 1.01 \\
HII 3179 & HD 24194 & 3:51:56.89 & 23:54:06.1 & $133.1_{-1.1}^{+1.1}$ & $5845_{-172}^{+144}$ & $1.42_{-0.02}^{+0.02}$ & 1.10 & 1.00 
\enddata
\medskip
\tablenotemark{a}{Distances from {\em Gaia} DR2 catalog parallaxes \citep{GAIA2018} 
using the method of \citet{Bailer-Jones+2018}.} 
\tablenotetext{b}{Stellar luminosities and effective temperatures from the {\em Gaia} DR2 catalog \citep{GAIA2018}. 
\tablenotetext{c}{24~$\mu$m excess ratio reported by \citet{Sierchio+2010} 
except for HII~1776 that is reported by \citet{Carpenter+2009}; significant detections are indicated in bold.} }
\end{deluxetable}

\subsection{ALMA Parameters and Observations}
The ALMA Band 6 (1.3~mm) observations were performed in 7 execution blocks 
in the period from 2019 December 12 to 2019 December 18 with 
42 to 44 available antennas (program 2019.1.00251.S). 
The array was in configuration C43-2, which provided a beam size of 
$\sim1\farcs5$ (200~au) and $12''$ (1600~au) maximum recoverable scale. 
Each star was observed for 4.2 minutes. 
The correlator setup included three spectral windows centered at 232.5, 247.0, and 245.0 GHz 
with 2.0~GHz bandwidth and 128 channels (TDM mode), 
and one spectral window centered at 230.538~GHz with 1.875~GHz bandwidth 
and 3840 channels (FDM mode). 
The field of view was set by the  $27''$ (FWHM) primary beam of the ALMA antennas at 1.3~mm. 
The datasets were processed by the ALMA standard calibration pipeline using the 
Common Astronomy Software Applications ($\mathtt{CASA}$) software package (version 5.6.1) \citep{McMullin+2007}.
Table~\ref{tab:observationslog} presents a log of the ALMA observations.
In brief, the time dependence of phase and amplitude were calibrated with frequent observations 
of the nearby  quasar J0357+2319, the passband response was calibrated with observations of the 
strong source J04323-0120, 
and the absolute flux scale was referenced to ALMA observations of Solar System objects 
with an estimated accuracy of 10\%.

\begin{deluxetable}{lccccc}
\tablecaption{ALMA Observations Log
\label{tab:observationslog}
}
\tabletypesize{\scriptsize}
\tablehead{
\colhead{Date}  & \colhead{Antennas}  & \colhead{Baselines} & \colhead{pwv} & \colhead{Passband} & \colhead{Phase}  \\
\colhead{}    & \colhead{}               & \colhead{(m)}       & \colhead{(mm)} & \colhead{}        & \colhead{}
}
\startdata
2019 December 12 & 43 & 15.0 - 313.7 & 1.0 & J0423-0120 & J0357+2319 \\
2019 December 13 & 43 & 15.0 - 313.7 & 1.2 & J0423-0120 & J0357+2319 \\
2019 December 14 & 44 & 15.0 - 313.7 & 2.8 & J0423-0120 & J0357+2319 \\
2019 December 15 & 42 & 15.1 - 313.7 & 0.8 & J0423-0120 & J0357+2319 \\
2019 December 16 & 43 & 15.1 - 313.7 & 0.8 & J0423-0120 & J0357+2319 \\
2019 December 17 & 43 & 15.1 - 312.7 & 1.1 & J0423-0120 & J0357+2319 \\
2019 December 18 & 42 & 15.1 - 312.7 & 1.0 & J0423-0120 & J0357+2319 \\
\enddata
\end{deluxetable}

\subsection{ALMA Sensitivity in Context}
\begin{figure}
    \centering
    \includegraphics[width=0.75\linewidth]{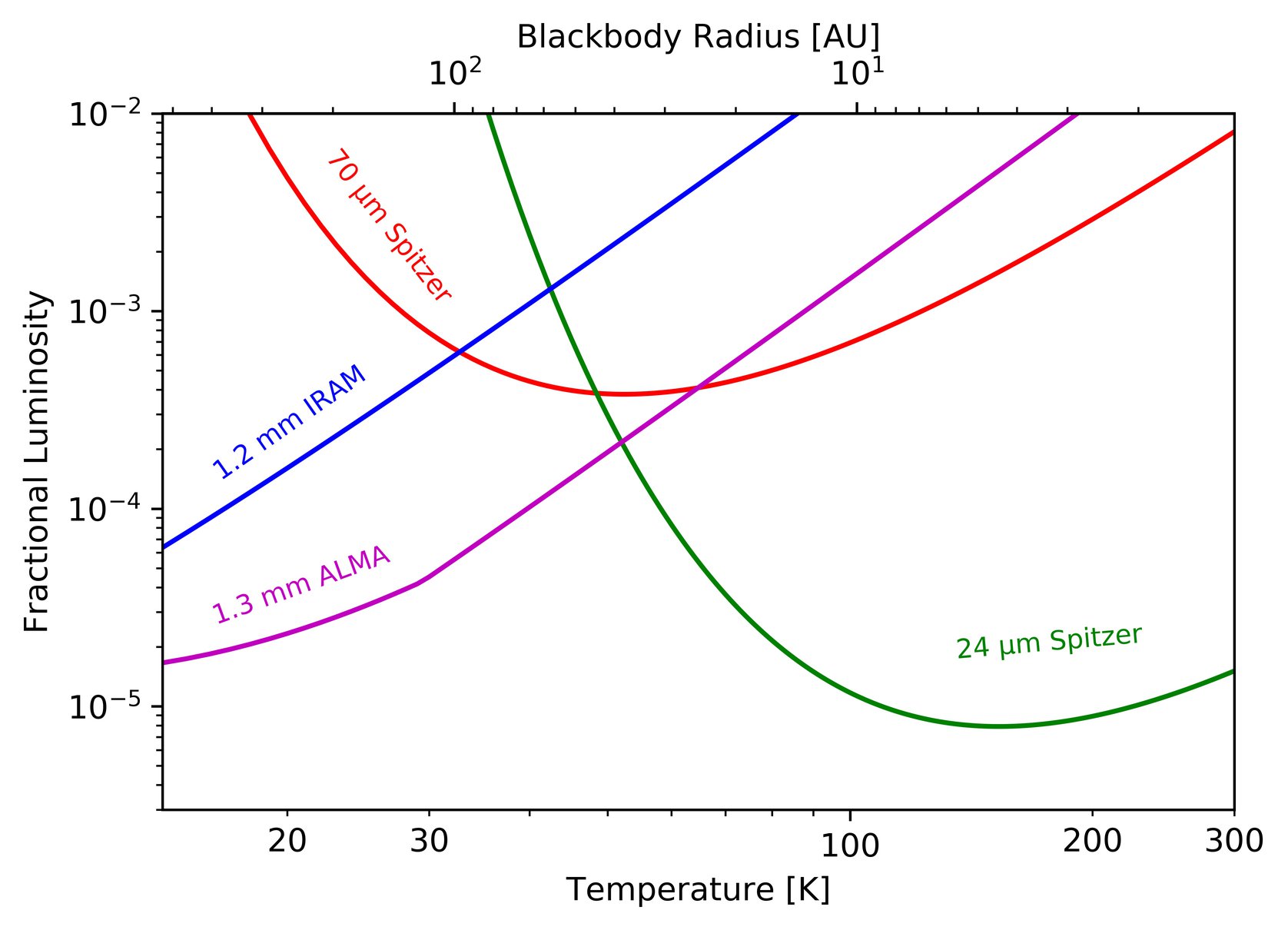}
    \caption{Detection limits for various debris disk surveys of Solar-type stars in the Pleiades, 
    in terms of fractional luminosity ($L_{dust}/L_{star}$).
    The colored curves show the typical limits from the {\em Spitzer} 24 $\mu$m surveys of the Pleiades 
    sample targeted for ALMA \citep[green,][]{Sierchio+2010}, 
    the {\em Spitzer} 70 $\mu$m observations of a subsample of 20 systems \citep[red,][]{Stauffer+2005},
    and IRAM 30 meter 1.2 mm observations of 5 of those systems \citep[blue,][]{Roccatagliata+2009}.
    The purple line shows the ALMA detection limit, which improves on the millimeter sensitivity by 
    about an order of magnitude (for more than an order of magnitude more systems).}
    \label{fig:sensitivity}
\end{figure}

A useful way to contextualize the sensitivity of these ALMA observations 
is to compare detection limits from surveys at different wavelengths
in terms of fractional luminosity, which can be expressed 
for blackbody emission simply as a  function of dust temperature, stellar effective temperature, 
and dust-to-star flux ratio \citep[e.g.][]{Beichman+2006,Greaves+2009}.
The ring-like appearance of most debris disks supports the assumption of a single value 
for the dust temperature as a reasonable approximation, 
even as some debris disks sometimes do span a broad range of radii \citep[e.g.][]{Kalas+2006}.
The fractional luminosity provides a proxy for the total cross-sectional area of dust.
Figure~\ref{fig:sensitivity} shows curves for the minimum detectable values of 
$L_{dust}/L_{star}$ for a typical Solar-type star in the Pleiades sample
($L_{star} = 1.8~L_{\odot}, T_{eff} = 5825$~K) as a function of $T_{dust}$ for 
(1) the Pleiades sample observed at 24 $\mu$m by {\em Spitzer} with flux density 
more than 10\% in excess of the extrapolated photospheric flux density
(32\% detected, see \citet{Sierchio+2010}), 
(2) the subsample of 20 systems observed at 70 $\mu$m by {\em Spitzer} with $3 \times rms$ = 21 mJy (no detections, \citet{Stauffer+2005}),
(3) the small group of 5 systems observed from the 70 $\mu$m subsample at 1.2 mm with the IRAM 30 meter telescope
with $3 \times rms$ = 2.1 mJy,  which are the most sensitive millimeter observations to date (no detections, \citet{Roccatagliata+2009}),
and (4) the 1.3~mm ALMA observations for $3 \times rms$ = 160 $\mu$Jy~beam$^{-1}$,
representative of the Pleiades sample (see Table~\ref{tab:results},
where the rms value for each target was determined by the noise statistics of the entire image),
as well as an additional fall off as the largest belts become spatially resolved (see \S\ref{sec:limits}). 
The millimeter curves also account for steeper than blackbody spectra at these long wavelengths 
(see \S\ref{sec:emissivity} for details).
The ALMA detection limit expressed in fractional luminosity is $L_{dust}/L_{star} = 2\times10^{-4}$ 
for $T_{dust} = 50 $~K. This limit is comparable to the {\em Spitzer} 24 $\mu$m threshold for similar values 
of $T_{dust}$, and substantially better for dust belts at lower temperatures.

\subsection{Background Source Confusion}
\label{sec:background}
The Pleiades cluster shows a reflection nebula from a nearby dust cloud, and the presence of this 
dusty cloud has the potential to confuse searches for excess emission from stars in the cluster.  
For ALMA, the interferometer response strongly spatially filters extended emission, 
so confusion from this cloud is unlikely to be problematic. 
Moreover, with a $1\farcs5$ beam, extragalactic millimeter source counts show that chance 
alignments with background contaminants are insignificant \citep[e.g.][]{Carniani+2015,Gonzalez-Lopez+20}.
The larger {\em Spitzer} beams at 24 $\mu$m ($6''$) and 70 $\mu$m ($20''$) clearly encompass any 
circumstellar dust 
emission around Pleiades members but also raise the spectre of confusion with background sources. 
\citet{Stauffer+2005} suggest that up to 2 out of 5 excess detections at 24 $\mu$m in their sample of 
20 stars could be due to confusion with background Active Galactic Nuclei or Galactic cirrus.

\section{Results} 
\label{sec:results}

\begin{figure}
    \centering
    \includegraphics[width=\linewidth]{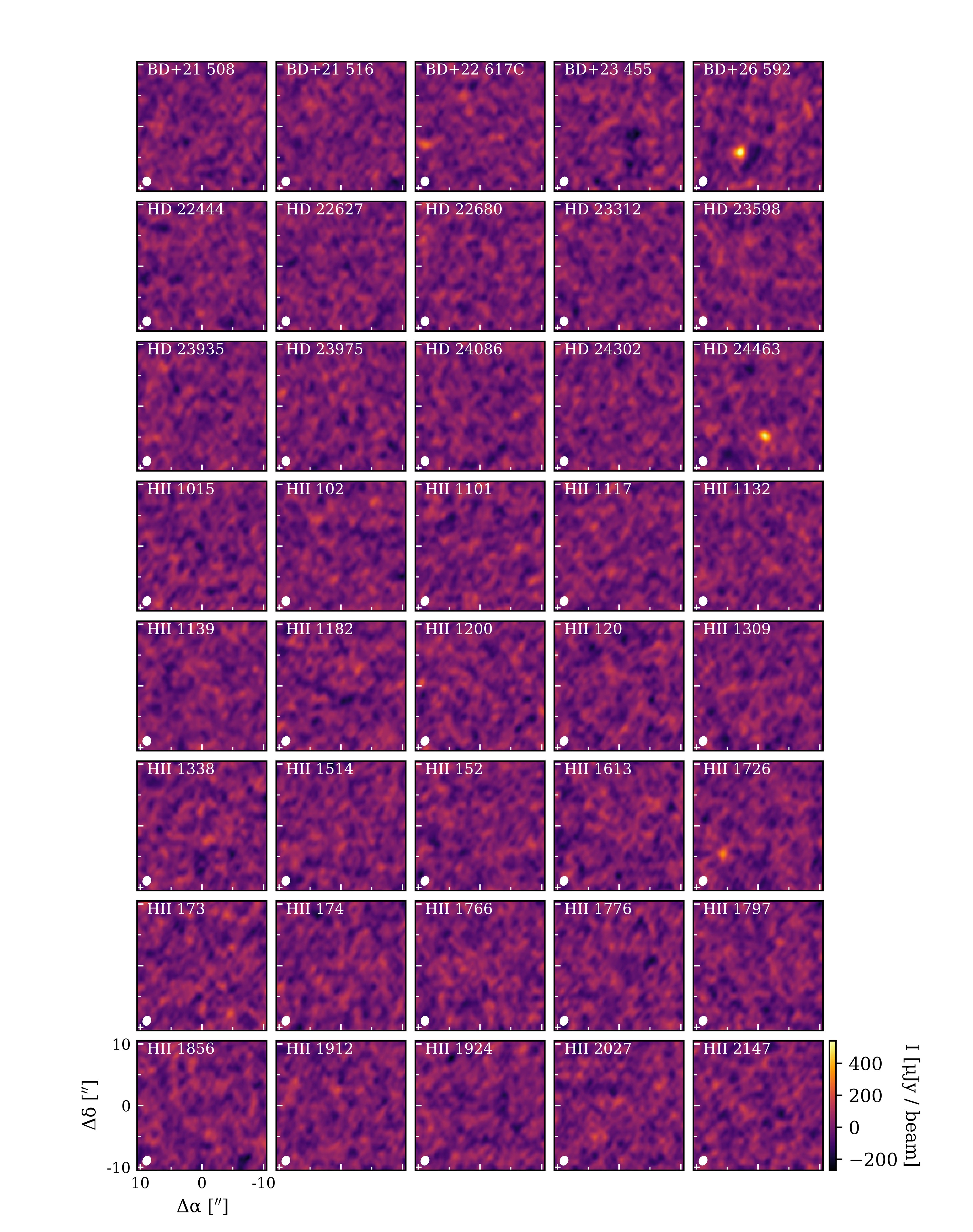}
    \caption{
    ALMA 1.3~mm images of the Pleiades FGK star sample, with typical 
    at the origin of its panel.  The synthesized beam 
    is indicated by the ellipse in the lower left corner.  
    No circumstellar dust emission is detected from any of the target stars.
    } 
    \label{fig:imagegallery1}
\end{figure}

\begin{figure}
    \centering
    \includegraphics[width=\linewidth]{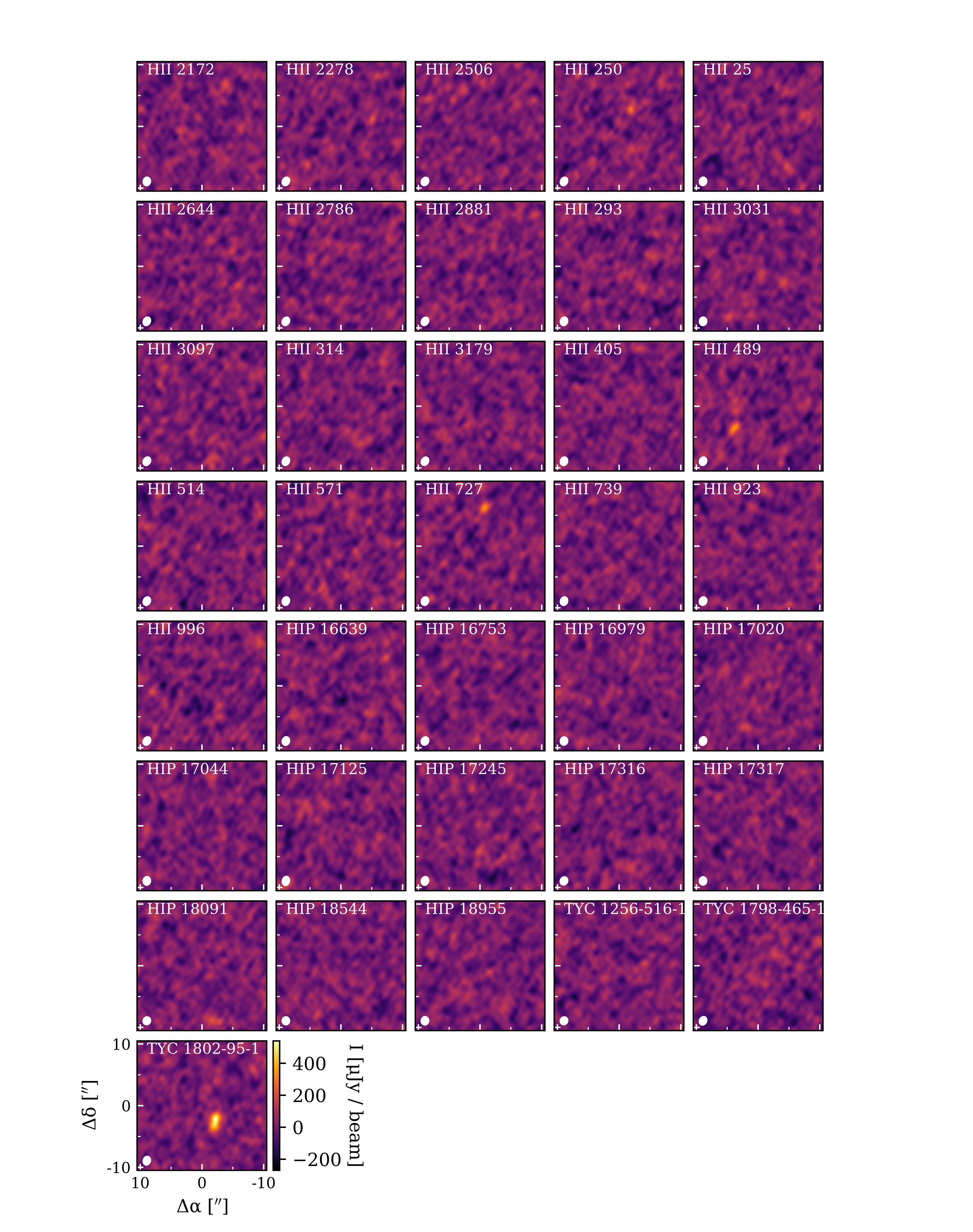}
     \caption{
    ALMA 1.3~mm images of the Pleiades FGK star sample (continued). 
    }
    \label{fig:imagegallery2}
\end{figure}

We used the {\tt tclean} task in CASA to generate 
continuum images for each star in the sample. We adopted natural weighting 
to obtain the best point source sensitivity, iterated to a threshold of 
$2 \times$ the rms noise level, and applied a primary beam correction.
Table~\ref{tab:results} lists the beam size and rms noise for 
the image for each star, which ranges from 48 to 73 $\mu$Jy~beam$^{-1}$.
The median rms noise was 54~$\mu$Jy~beam$^{-1}$.
Figures~\ref{fig:imagegallery1} and~\ref{fig:imagegallery2} 
shows galleries of ALMA 1.3~mm images centered on each 
of the target stars in the full Pleiades sample. 

\startlongtable
\begin{deluxetable}{lcrc}
\tablecaption{ALMA 1.3 millimeter Imaging
\label{tab:results}
}
\tablewidth{0pt}
\tabletypesize{\scriptsize}
\tablehead{
\colhead{Name} & \colhead{Beam FWHM} & \colhead{P.A.} & \colhead{rms noise}\\
\colhead{} & \colhead{(arcsec)} & \colhead{(deg)} & \colhead{($\mu$Jy~beam$^{-1}$)}}
\startdata
HIP 17317 & $1.56 \times 1.42$ & $ 0.8$ & 48 \\
BD+21 508 & $1.56 \times 1.42$ & $ 1.4$  & 50 \\
HD 23312 & $1.55 \times 1.42$ &  $2.5$  & 50 \\
HD 24463* & $1.62 \times 1.41$ &  $10.6$  & 58 \\
HD 22627 & $1.60 \times 1.41$ &  $-6.6$  & 51 \\
HIP 17044 & $1.60 \times 1.41$ & $ -7.7$  & 51 \\
HD 23598 & $1.63 \times 1.40$ &  $-0.8$  & 53 \\
HD 24086 & $1.63 \times 1.40$ &  $0.3$  & 51 \\
HD 23975 & $1.62 \times 1.40$ &  $-0.2$  & 51 \\
HD 22444 & $1.56 \times 1.42$ &  $-12.7$  & 51 \\
TYC 1798-465-1 & $1.64 \times 1.38$ &  $-26.6$  & 57 \\
HIP 16979 & $1.56 \times 1.42$ &  $-14.4$  & 49 \\
HD 22680 & $1.58 \times 1.42$ &  $-12.5$  & 52 \\
HIP 16639 & $1.62 \times 1.40$ & $ -18.9$  & 55 \\
HII 25 & $1.61 \times 1.40$ &  $-12.3$  & 54 \\
HII 102 & $1.58 \times 1.42$ &  $-11.1$  & 52 \\
HII 1132 & $1.57 \times 1.43$ &  $-9.1$  & 50 \\
HII 1139* & $1.58 \times 1.42$ & $ -9.1$  & 55 \\
HII 1766 & $1.63 \times 1.39$ &  $-8.2$  & 54 \\
HII 2172 & $1.61 \times 1.40$ &  $-9.0$  & 52 \\
HII 3031 & $1.60 \times 1.41$ &  $-6.5$  & 52 \\
TYC 1802-95-1* & $1.65 \times 1.38$ &  $-17.2$  & 59 \\
HIP 17020 & $1.64 \times 1.39$ &  $-19.5$  & 50 \\
HIP 17245 & $1.66 \times 1.39$ &  $-16.7$  & 54 \\
HIP 17125 & $1.71 \times 1.37$ &  $-12.5$  & 55 \\
BD+21 516 & $1.56 \times 1.42$ &  $-18.8$  & 49 \\
HIP 18544 & $1.52 \times 1.43$ &  $-3.1$  & 50 \\
BD+23 455 & $1.64 \times 1.37$ &  $-27.2$  & 54 \\
HIP 16753 & $1.61 \times 1.40$ &  $-25.6$  & 52 \\
BD+26 592* & $1.68 \times 1.38$ & $ -15.7$  & 61 \\
TYC 1256-516-1 & $1.52 \times 1.41$ &  $-30.1$  & 51 \\
HIP 18091* & $1.51 \times 1.42$ &  $-27.3$  & 73 \\
HD 23935 & $1.65 \times 1.39$ &  $-14.9$  & 52 \\
BD+22 617C & $1.58 \times 1.43$ &  $-7.6$  & 50 \\
HIP 18955 & $1.57 \times 1.43$ &  $-5.1$  & 53 \\
HIP 17316 & $1.57 \times 1.40$ &  $-27.6$  & 52 \\
HD 24302 & $1.63 \times 1.40$ &  $-17.0$  & 49 \\
HII 293 & $1.64 \times 1.38$ &  $-22.2$  & 58 \\
HII 405 & $1.63 \times 1.38$ &  $-21.1$  & 53 \\
HII 489 & $1.63 \times 1.38$ &  $-22.2$  & 54 \\
HII 571 & $1.64 \times 1.38$ &  $-20.5$  & 57 \\
HII 727 & $1.63 \times 1.38$ &  $-22.6$  & 56 \\
HII 739 & $1.64 \times 1.38$ &  $-21.9$  & 54 \\
HII 923 & $1.61 \times 1.41$ &  $-24.9$  & 53 \\
HII 996 & $1.66 \times 1.35$ &  $-29.8$  & 60 \\
HII 1117 & $1.62 \times 1.38$ &  $-25.0$  & 52 \\
HII 1309 & $1.63 \times 1.38$ &  $-23.8$  & 55 \\
HII 1338 & $1.63 \times 1.38$ &  $-24.7$  & 54 \\
HII 1514 & $1.63 \times 1.38$ &  $-24.2$  & 55 \\
HII 1613 & $1.62 \times 1.38$ &  $-25.4$  & 55 \\
HII 1726 & $1.63 \times 1.37$ &  $-25.0$  & 56 \\
HII 1797 & $1.62 \times 1.38$ &  $-26.6$  & 52 \\
HII 1856 & $1.63 \times 1.38$ &  $-25.8$  & 55 \\
HII 1912 & $1.63 \times 1.37$ &  $-25.6$  & 52 \\
HII 1924 & $1.62 \times 1.39$ &  $-28.1$  & 53 \\
HII 2027 & $1.63 \times 1.37$ &  $-26.0$  & 54 \\
HII 120 & $1.63 \times 1.35$ &  $-31.4$  & 56 \\
HII 152 & $1.63 \times 1.36$ &  $-31.9$ & 53 \\
HII 173 & $1.66 \times 1.36$ &  $-25.8$  & 58 \\
HII 174 & $1.66 \times 1.35$ &  $-27.9$  & 56 \\
HII 250 & $1.65 \times 1.35$ &  $-28.3$  & 55 \\
HII 314 & $1.65 \times 1.35$ &  $-28.4$  & 55 \\
HII 514 & $1.67 \times 1.35$ &  $-28.4$  & 57 \\
HII 1015 & $1.66 \times 1.35$ &  $-27.9$  & 53 \\
HII 1101 & $1.65 \times 1.35$ &  $-28.2$  & 57 \\
HII 1182 & $1.64 \times 1.39$ &  $-33.0$  & 56 \\
HII 1200 & $1.64 \times 1.35$ &  $-35.0$  & 56 \\
HII 1776 & $1.66 \times 1.35$ &  $-29.0$  & 55 \\
HII 2147 & $1.64 \times 1.35$ &  $-32.3$  & 56 \\
HII 2278 & $1.67 \times 1.35$ &  $-29.7$  & 55 \\
HII 2506 & $1.64 \times 1.36$ &  $-33.6$  & 55 \\
HII 2644 & $1.66 \times 1.35$ &  $-29.9$  & 52 \\
HII 2786 & $1.65 \times 1.34$ &  $-31.9$  & 57 \\
HII 2881 & $1.65 \times 1.34$ &  $-32.2$  & 53 \\
HII 3097 & $1.67 \times 1.35$ &  $-29.9$  & 54 \\
HII 3179 & $1.65 \times 1.34$ &  $-32.2$  & 54 \\
\enddata
\tablecomments{$^*$ Serendipitous emission source in the field of view\\}
\end{deluxetable}

\subsection{Pleiades 1.3~mm Dust Continuum Emission Limits}
\label{sec:limits}
We visually inspected the images and did not identify any significant sources of emission at the target star 
locations. We consider the ALMA detection limit to be $3\times$ the rms noise level, an appropriate metric
given that we have accurate {\em a priori} knowledge of the stellar positions. 
As an additional check, we fitted a point source model to the visibilities for each target star using the 
{\tt UVMultiFit} library \citep{Marti-Vidal+2014}, restricting the fit source location to within 
$0\farcs2$ offsets of the stellar position; this procedure also did not reveal any significant sources
of emission. 

We can translate the ALMA upper limits into estimates of dust mass by making standard assumptions,  
i.e. $ M_{dust} = F_\nu d^2/(\kappa_\nu B_\nu(T_{dust}) ) 
\approx 0.012~(F_{1.3~mm}/54~\mu Jy) (d/133.5~pc)^2 (T_{dust}/40~K)~M_{\oplus} $
where $F_{1.3~mm}$ is the 1.3~mm flux density, 
$\kappa_{1.3~mm} = 2.0$~cm$^2$~g$^{-1}$ is the dust mass opacity \citep{Beckwith+1990}, 
and $B_\nu(T_{dust})$ is the Planck function taken to be in the Rayleigh-Jeans regime.
We have adopted the dust mass opacity for circumstellar disks advocated by \citet{Beckwith+1990} 
in order to facilitate comparisons with other studies that make this common assumption.
With this assumption, taking $3\times$ the median rms noise level of the sample 
and $M_{dust} = 40$~K leads to a dust mass limit of 0.036~M$_{\oplus}$ (about $3\times$ the mass of the Moon).
We also made a stacked image from the 76 continuum images, to obtain a deeper constraint 
on  the average flux density of the undetected sources. This stacked image shows 
no significant emission at the center and has an rms noise level of 6.1~$\mu$Jy~beam$^{-1}$. 
Adopting the same assumptions leads to a dust mass limit of 0.004~M$_{\oplus}$
for the mean level of dust around stars in this sample.

The typical FWHM beam size of $1\farcs5$ (200~au) comfortably encompasses dust belts with radii up to 100~au.
For larger dust belts, the emission spreads across multiple beams, and the sensitivity for detection is 
generally lower, depending in detail on the radius, width and viewing geometry. 
To quantify the dust belt detection threshold as a function of size in a simple way, we made a series of 
natural weight images with Gaussian tapers
that increased in $0\farcs5$ steps from 0 to $12\farcs5$. 
This process produces a sequence of larger beams to cover larger dust belts 
(at any viewing geometry) at the cost of poorer point source sensitivity. 
This sensitivity as a function of size scale has a similar shape for all  
of the targets because the  $u,v$ plane coverage of the ALMA observation is very similar.
The sensitivity decreases by a factor of 2 at a beam size 
of $4\farcs4$ and a factor of 3 at $6\farcs9$. 
While there is some dispersion in the noise levels at the largest tapers among the different targets, 
the increase in rms as a function of $\theta$, the beam size FWHM (arcsec),
is described well by the expression $1.000 + 0.32(\theta - 1.5) + 0.010(\theta - 1.5)^2$
for beam sizes in the range of interest.
We will adopt this dependence of sensitivity with size scale in 
comparing the observations to population synthesis models in \S\ref{sec:models}.

\subsection{Serendipitous Source Detections}

\begin{figure}
    \centering
    \includegraphics[width=0.95\linewidth]{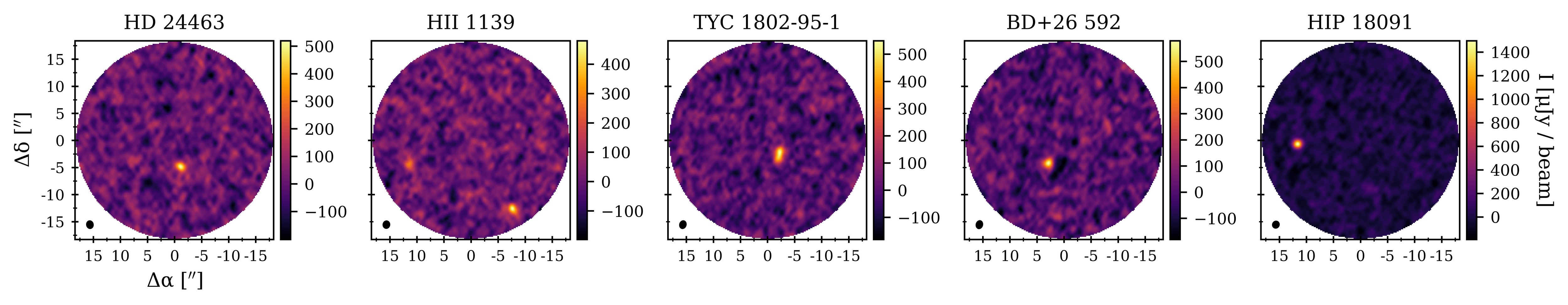}
    \caption{ALMA 1.3 mm continuum images for Pleiades fields where significant ($>5\sigma$) 
    emission sources were detected
    that are not associated with the target stars located at the field center. Table~\ref{tab:detections} lists the 
    locations and flux densities of these sources. The beam sizes are indicated in the lower left corner of each panel.
    }
    \label{fig:serendipitous_sources}
\end{figure}

For five targets, we identified significant 
($>5\sigma$) 
sources of emission within the field of view but not associated 
with the centrally located star (HD 24463, HII 1139, TYC 1802-95-1, BD+26 592, and HIP 18091). 
For these serendipitous sources, we used $\mathtt{UVMultiFit}$ to fit a Gaussian model to the 
detected source(s), simultaneous with a central point source. Table~\ref{tab:detections} lists fitted 
offsets (east and north) from the field center, absolute coordinates,
and flux densities.  Figure~\ref{fig:serendipitous_sources} shows the images that contain these sources. 
The source in the field of TYC~1802-95-1 is clearly resolved.
We checked the 2MASS infrared K band images of all of these sources 
for near-infrared counterparts  and did not  find any, although blending 
made the relevant region difficult to assess in some cases. 
The lack of any obvious near-infrared counterparts suggests the emission is not associated 
with stars located in the Pleiades cluster.
Most likely they are dusty galaxies in the background.  
According to the 1.2~mm source counts of \citet{Gonzalez-Lopez+20}, 
approximately 5 sources exceeding $\gtrsim500~\mu$Jy are expected across the sample 
($\sim5\sigma$ at the primary beam half power width), consistent with these detections.

\begin{deluxetable}{lcccc}
\tablecaption{Serendipitous Source Detections\tablenotemark{a}
\label{tab:detections}
}
\tablewidth{0pt}
\tabletypesize{\scriptsize}
\tablehead{
\colhead{Target Field} &  \colhead{offset} & \colhead{RA} & \colhead{Dec} 
& \colhead{Flux Density} \\
\colhead{} & \colhead{(E$''$, N$''$)} & \colhead{(J2000)} & \colhead{(J2000)} & \colhead{(mJy)} \\
}
\startdata
HD 24463  & -1.1, -5.6 & 3:54:21.52 & 24:04:26.0 & 1.48$\pm$0.22 \\
HII 1139  & -8.6, -13.1 & 3:46:39.43 & 23:06:23.2 & 1.84$\pm$0.31 \\
HII 1139  & 11.1, -4.4 & 3:46:40.74 & 23:06:31.9 & 1.38$\pm$0.22 \\
TYC 1802-95-1  & -2.2, -2.3 & 3:34:47.05 & 26:05:38.0 & 1.25$\pm$0.15 \\
BD+26 592  & 3.0, -4.1 & 3:39:53.90 & 26:42:56.2 & 0.95$\pm$0.11 \\
HIP 18091  & 11.7, -0.6 & 3:52:01.57 & 19:35:47.2 & 2.58$\pm$0.16 \\
\enddata
\tablenotetext{a}{Positions and flux densities obtained from Gaussian model fits to the visibilities and corrected for primary beam attenuation. The positional uncertainties are typically $<0\farcs2$ ($1\sigma$) in each coordinate.}
\end{deluxetable}

\section{Discussion} 
\label{sec:discussion}

The ALMA 1.3~mm survey of a sample of FGK type stars in the 115~Myr-old Pleiades cluster 
did not  result in any detections of circumstellar dust emission. 
These data provide new limits on dust belt fractional luminosities and masses in a regime 
of low temperatures and large radii that significantly improve on previous observations 
of Solar-type stars of Pleiades age. In particular, the dust mass limit derived from 
stacking the non-detections is $<0.004$~M$_{\oplus}$, about two orders of magnitude 
lower than from previous millimeter observations of such stars in the Pleiades 
(for the same dust opacities and temperatures; 
\citealt{Greaves+2009,Roccatagliata+2009}).
While a handful of Solar analogs believed to be at an age similar to the Pleiades are known 
to harbor massive ($\gtrsim0.10$~M$_{\oplus}$) dust belts at Kuiper Belt scales and beyond
\citep{Holland+2017},
such systems are evidently rare and not found in this Pleiades sample.
For example, the disk around the nearby $\sim80-200$~Myr-old G2 star HD~107146 
would be 500~$\mu$Jy at 1.3~mm at Pleiades distance \citep{Williams+2004,Ricci+2015} 
and readily detected in this survey. 
We examine these Pleiades survey results in the context of debris disk evolution models 
based on collisional evolution. 

\subsection{Limits on Fractional Luminosity}
As Figure~\ref{fig:sensitivity} shows, these ALMA observations provide new sensitivity 
to dust belts with $T \lesssim 40$ K and fractional luminosity $> 10^{-4}$.
For warm disks with blackbody radii $\sim5$~au, the fractional luminosity limit implied by the observations
is $f \sim 5 \times 10^{-3}$. For colder disks with blackbody radii $\sim100$~au, the fractional luminosity limit 
is $f \sim 5 \times 10^{-5}$.
None of the Pleiades stars with significant 24~$\mu$m excess emission are detected at 1.3 mm. 
Given that these systems are also not detected by {\em Spitzer} at the shorter wavelength of 8~$\mu$m, 
the circumstellar dust must be at intermediate temperatures, and therefore 
concentrated at intermediate radii (from a few au to a few 10's of au).
The previous non-detections of Pleiades targets at 70~$\mu$m and 1.2~mm are all consistent with the 
ALMA results (e.g., see Figure.~\ref{fig:sensitivity}).  
The lack of 1.3~mm dust continuum emission detections implies an absence of massive and 
cold debris belts at large radial distances around the target stars. 

\subsection{Collisional Evolution Models}
\label{sec:models}
Multi-wavelength observations of stars of different ages and spectral types provide constraints on 
debris disk incidence and fractional luminosity that have led to the development of various models to 
describe debris disk evolution based on collisions and ``stirring'' mechanisms that dynamically 
excite  planetesimals leftover from the planet formation process 
\citep[e.g.,][]{WyattDent2002,DominikDecin2003, KenyonBromley2004a,Wyatt+2007a}. 

\citet{Najita+2022} used new coagulation models to assess the possibility that 
the observed population of cold debris disks around Sun-like stars evolve from the 
$\sim25$\% of T~Tauri systems that harbor large disks with ringed substructures. 
They show that the resulting ``bright stalwart'' debris disks by themselves can 
account for the available constraints on cold debris disk incidence and luminosities. 
In particular, models that start with  rings at radii of 45 or 75~au with at least 
a few Earth masses in solids, using intermediate planetesimal formation efficiencies, 
readily produce cold belts with $L_{disk}/L_{star} \sim10^{-3}$ at 10-100~Myr that 
subsequently fade.  In this scenario, one would expect the ALMA 1.3~mm survey of 
the Pleiades to reveal many such belts. Since none were detected, the descendents 
of the large and ringed protoplanetary disks, if present, must be at least an 
order of magnitude less luminous at Pleiades age. These limits suggest different initial conditions 
or evolutionary pathways. 

The analytical evolutionary model of \citet{Wyatt+2007a}, building on \citet{WyattDent2002} 
and \citet{DominikDecin2003}, provides a practical framework for comparison with observations. 
In brief, this model assumes steady-state evolution in a catastrophic collisional cascade.
Every star is assumed to be born with a planetesimal belt with radius $r$, narrow width $dr$, 
mass $M_{disk}$, and a size distribution of colliding objects $n(D) \propto D^{2-3q}$, 
where $q = 11/6$ \citep[i.e. collisional equilibrium,][]{Dohnanyi1969, Tanaka+1996}.
In this analytical model, the belt properties are an initial condition; 
no physical explanation is provided for the origin of the belts. 
The production of dust is a top-down process that requires fragmentation of larger belt objects,
and the evolution on long timescales is therefore determined by the collisional lifetime of the largest planetesimals,
$t_c$ (see equation 7 of \citet{Wyatt+2007a}).
Then the time evolution of the disk mass is determined by the differential equation 
$dM_{tot}/dt = -M_{tot}/t_c$, which implies $M_{tot}(t) = M_{tot}(0) / [1+t/t_c(0)]$,
i.e. the debris disk mass (and fractional luminosity) decrease as $t^{-1}$.
This is an approximation, as catastrophic collisions gradually reduce the size of the largest 
planetesimals over time, which results in a shorter collisional timescale and a faster decrease 
of disk mass \citep{KenyonBromley2016}.
Nonetheless, this basic model has been successful in reproducing many observations of debris disks, 
including 
the properties of debris disks around A type stars of various ages at 24~$\mu$m and 70~$\mu$m \citep{Wyatt+2007b}, 
the incidence and excess emission of nearby FGK type stars at 24, 70, 100 and 160 $\mu$m \citep{Kains+2011,Sibthorpe+2018}, 
and the (low) detection rate of debris disks around M stars \citep{MoreyLestrade2014}.
Additional intuition about the time evolution behavior may be obtained by noting that
$t_c \propto r^{3.5}/M_{tot}$ in this model, 
or $t_c \propto 1/(M_{tot}T^7)$ using blackbody radiative equilibrium  
to relate dust temperature to belt radius, i.e. $T\propto L_{star}^{1/4} r^{-1/2}$. 
The extremely steep dependence of timescale on dust temperature 
implies that warm (small radius) belts evolve much more rapidly than cold (large radius) belts.

\subsubsection{Steady-State Evolution Parameters}
We applied the collisional evolution model as described by \citet{Wyatt+2007b} 
to the population of FGK stars in Pleiades  observed by ALMA. 
The initial disk parameters  
are the disk mass $M_{tot}(0)$, 
the disk radius $r$ and width $dr$, 
the material density $\rho$, 
maximum planetesimal diameter $D_c$, 
minimum dust grain diameter, $D_{bl}$ 
and power-law index of the planetesimal size distribution, $q$.
The time evolution is determined by the 
dispersal threshold $Q_{D}^{*}$, 
eccentricity $e$, and inclination $i$. 
We fix $dr = r/2$, $\rho = \textrm{2700 kg m}^{-3}$, $q = 11/6$, and $e/i = 1$,
following the analysis of nearby FGK stars by \citet{Kains+2011}.
The disk masses are drawn from a log-normal distribution with mean M$_{mid}$ and
and width 1.14 dex (as determined by \citealt{AndrewsWilliams2005}), and the disk radii 
are drawn from a power law distribution $N(r) \propto r^{\gamma}$ from 1 to 1000~au.
The five free parameters in the model are M$_{mid}$, $\gamma$, $Q_{D}^{*}$, $e$ and $D_c$,
and we set these to the values determined by \citet{Sibthorpe+2018} from 
fitting to {\em Herschel} observations of debris disks around FGK stars near the Sun, 
i.e. M$_{mid} = 2.1$~M$_{\oplus}$,  $\gamma = -1.7$,
$Q_{D}^{*} = 500$~J~kg$^{-1}$, $e=0.05$, and $D_c = 450$~km.
With these assumptions, we can test if the same evolutionary model 
is consistent with observations of the Pleiades. 

The relevant stellar parameters in the models are 
luminosity, $L_{star}$, mass, $M_{star}$,  and effective temperature, $T_{eff}$.
We adopted values for $T_{eff}$, $L_{star}$ and $M_{star}$ from Table~\ref{tab:targets}.
(Highly accurate values for the stellar masses are not critical as the range across the sample is modest 
and model dependencies are weak.) 

We omitted Pels 173 from our analysis for the same reasons it was excluded by \citet{Sierchio+2010}:
the reddening to this star suggests it lies behind the Pleiades cluster and is not a member.
We also omitted HII~1726 and HII~2278 because the {\em Gaia} DR2 catalog does not provide a set of
stellar parameters. 

\subsubsection{Modified Blackbody Emission}
\label{sec:emissivity}
To calculate the spectrum of each dust belt, we assumed modified blackbody emission and obtain 
the flux from  $F_\nu = (2.95\times 10^{10}) B_{\nu}(\lambda, T) f r^{2} d^{-2} X_{l}^{-1}$
where $F_\nu$ is the disk's flux in units of Janskys, $f$ is the fractional luminosity,
$r$ is the belt radius in au, $d$ is  distance to the star in parsecs, 
and $X_l$ is an additional factor included to account for the steeper fall-off due to departures from
blackbody emission at long wavelengths, $X_{l} = ( \lambda\ / \lambda_0)^{\beta}$, where $\lambda_0$ is
a turnover wavelength (adopted to be 0.21~mm) and $\beta$ is power-law index that reflects the 
dust emissivity citep[see Section 2.3 of ][]{Wyatt2008}. 
A value of $\beta=1$ is often assumed for debris disks
\citep[e.g.,][]{Holland+2017}. However, detailed 
observations of the millimeter spectral slopes of nearby debris disks suggest that $\beta=0.5$ 
provides a more appropriate empirical description \citep{Macgregor+2016}. 
An additional effect is that the blackbody radius determined from radiative equilibrium 
tends to underestimate the true dust belt radius since small grains do not emit efficiently 
at long wavelengths \citep[e.g.,][]{Booth+2013}. 
This dependence acts to increase ALMA sensitivity to dust belts of larger radii.

\subsubsection{Pleiades Sample Population Synthesis}
We took random draws from the distributions of disk masses and radii to generate a population of belts around 
each of the 73 stars considered in the sample. These disks were then evolved to the Pleiades age of 115 Myr 
following collisional evolution, to produce a synthetic dust belt associated with each star. Then fluxes 
in the relevant bands were calculated for each one.
This process was repeated 10,000 times to build up meaningful statistics.

\begin{figure}
    \centering
    \includegraphics[width=\linewidth]{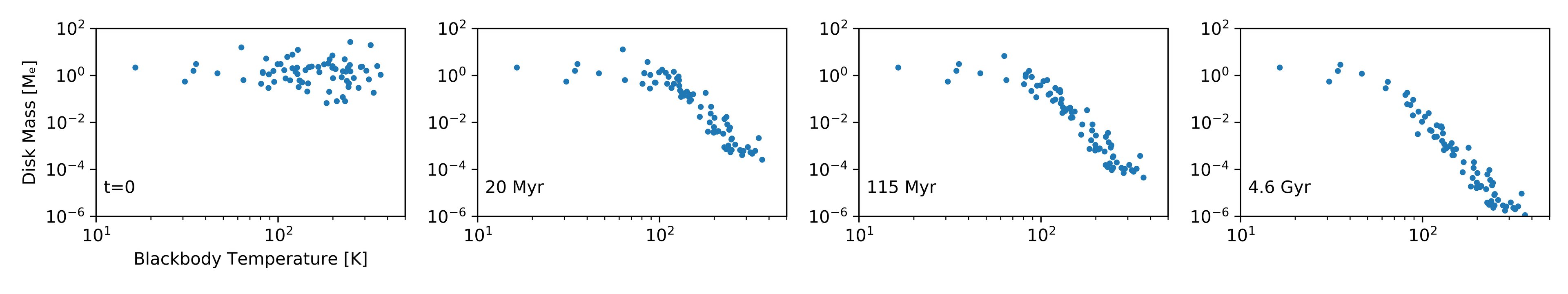}
    \includegraphics[width=\linewidth]{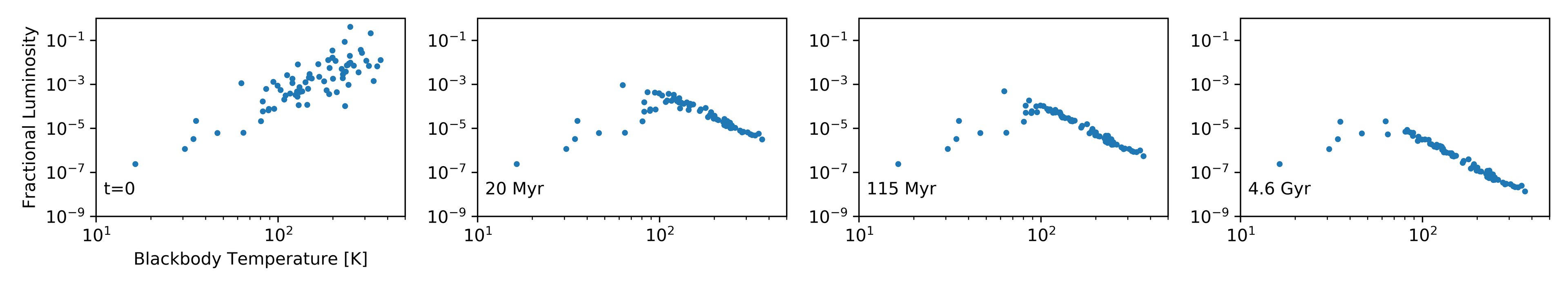}
    \caption{(top) Belt masses plotted against blackbody temperature for an example population of simulated debris disks 
    at four representative ages. The leftmost panel shows the initial masses, and subsequent panels show the 
    belt masses after collisional evolution at 20 Myr, 115 Myr, and 4.6 Gyr. By 20 Myr, the warmest belts are already significantly depleted, while the coldest belts experience negligible mass depletion. 
    (bottom) The fractional luminosities plotted against disk blackbody temperature for the same example population. 
    }
    \label{fig:beltevolution}
\end{figure}

Figure \ref{fig:beltevolution} illustrates the evolution of belt masses and fractional luminosities 
as a function of belt blackbody temperature for a single population of 73 stars at several ages.
The initial disk masses arise from sampling the log-normal distribution. 
By 20 Myr, the warmest belts already have significantly depleted mass, due to the steep 
dependence on dust temperature of the collisional lifetime of the largest planetesimals.
By contrast, the coldest belts ($T \lesssim 75 K)$ retain most of their mass at  20 Myr, 
at 115 Myr, and even on Gyr timescales. 
For a belt around a 1.0~M$_{\odot}$ star at 5 au radius, 
$t_c = 5.8 / (M_{tot} / M_{\oplus})$ Myr, while for a belt at 50 au radius,
$t_c = 8.8 \times 10^{4} / (M_{tot} / M_{\oplus})$ Myr and the initial mass is a dominant factor.
This behavior is also reflected in the fractional luminosities. 
At late times, the fractional luminosity is strongly dependent on blackbody temperature 
(or radius) as the belts are collisionally depleted. 
The coldest belts are governed by very long collisional lifetimes. 

\begin{figure}
    \centering
    \includegraphics[width=\linewidth]{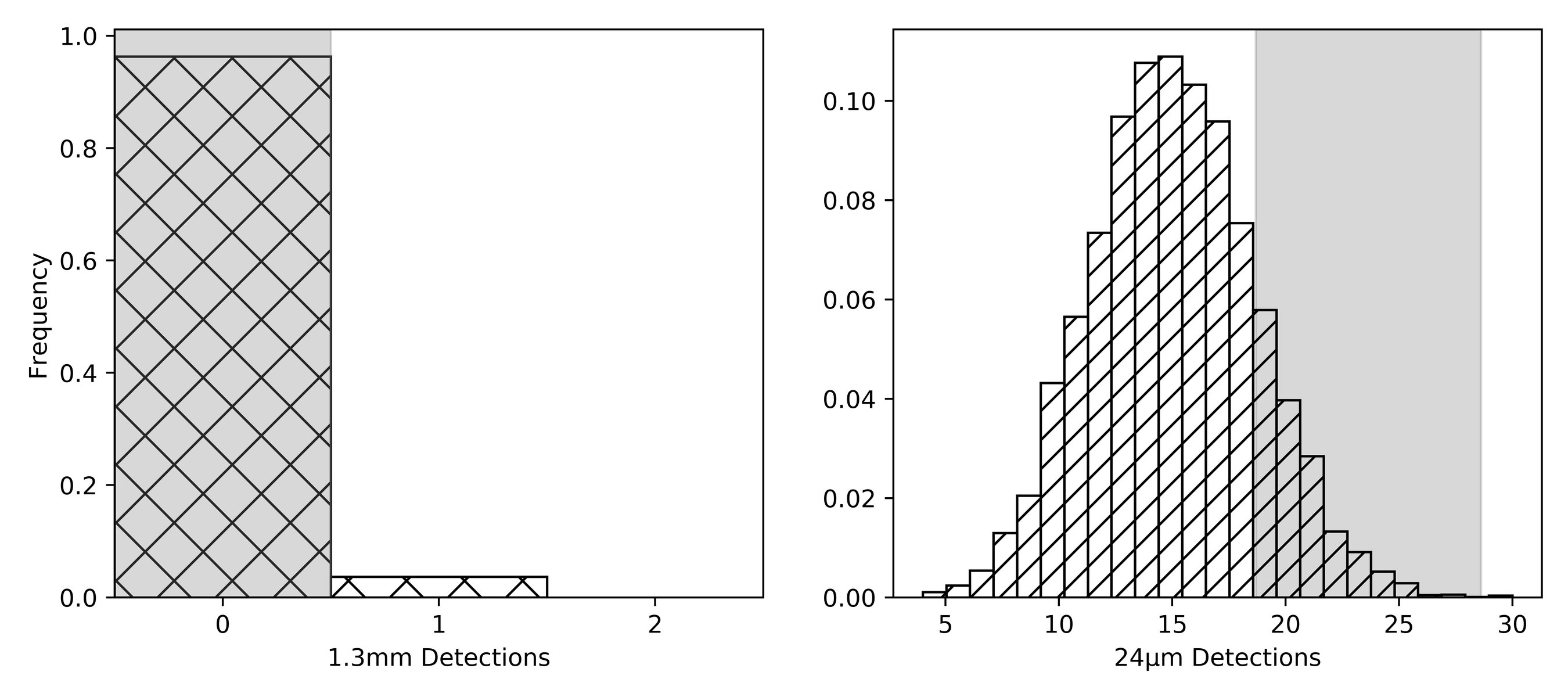}
 \caption{
 The hatched histograms show population synthesis models predictions for the detection frequency of debris disks
 in the Pleiades Solar-type star sample
 {\em (left)} by ALMA at 1.3~mm at the achieved sensitivity, 
 and {\em (right)} by {\em Spitzer} at $24~\mu$m, following the detection criterion of \citet{Sierchio+2010}.
 The grey shaded regions in each panel indicate the observed number (with associated uncertainties).
 }
 \label{fig:popmodelALMA+Spitzer}
\end{figure}

Figure \ref{fig:popmodelALMA+Spitzer} (left panel) shows the frequency of 1.3~mm detections 
of dust belts for the sample of 73 Pleiades stars, given the sensitivity achieved in the ALMA survey. 
From the 10,000 realizations of this Pleiades sample, there are 9670 with zero detections (96.7\%), 
325 with one detection (3.25\%), and 5 with 2 detections (0.05\%). 
Thus the ALMA nondetection of any significant disk emission from the Pleiades sample 
is in agreement with the most likely outcome of these models.
Much deeper integrations would be needed to detect a significant number of dust 
belts in this model population. 
Only about 5\% of the targets have 1.3~mm flux densities that exceed 20~$\mu$Jy, 
a level that would require almost an order of magnitude better sensitivity for detection 
(corresponding to $\sim100$ times longer integration times, or $\sim7$~hours on source with ALMA). 
This aspect of the modeled population is consistent with the flux density limit obtained
from stacking the individual ALMA images of Pleiades stars.

The assumed long wavelength slope of the millimeter emission, encapsulated in the parameter $\beta$, 
has a modest impact on the 1.3~mm emission and the resulting detection statistics. 
In particular, a shallower unmodified blackbody spectrum ($\beta=0$) 
extended (unrealistically) to the millimeter regime would result in at least one ALMA detection 
in 34\% of the simulated populations. For the steeper $\beta=1$, 
at least one ALMA detection from the sample would occur less than 1\% of the time. 

The population synthesis also provides predictions for 24~$\mu$m emission 
to compare with the {\em Spitzer} observations of the Pleiades sample. 
The 24~$\mu$m observations are much more sensitive than the 1.3~mm survey to the bulk of the dust belts.
\citet{Sierchio+2010} found significant excess emission 
from 23/71 stars, for a detection rate of  32$\pm$6.8\% of the sample  
(where an excess ratio $>1.1$ was considered significant, i.e. exceeding $3\sigma$).
Figure~\ref{fig:popmodelALMA+Spitzer} (right panel) shows the distribution of 
24~$\mu$m detections for the 10,000 simulated populations.
The mean of this distribution is 15.1, with standard deviation of 3.4,
or 21$\pm$4.7\% of the sample (taking into account that two targets were 
dropped because they lack {\em Gaia} catalog stellar properties). 
Given the dispersion in the model prediction and the observational uncertainties,
the population synthesis model predictions are consistent with the observations.
The trend of the data to the high side of the model predictions may 
indicate a contribution from background confusion. 
As noted in \S\ref{sec:background}, \citet{Stauffer+2005} cautioned that 10\% of 
their sample (2/20) may be contaminated by background 24~$\mu$m emission. 
If this same fraction were to hold for the full sample of \citet{Sierchio+2010}, 
then the number of stars with significant 24~$\mu$m {\em dust belt} detections 
would be lower, conceivably 16/71, close to the population synthesis prediction. 
However, the difference between the nominal number of 24~$\mu$m detections and the
model prediction is not statistically significant and should not be overinterpreted.

At 24~$\mu$m, the photospheric excess ratios predicted by the population synthesis models 
are always less than $2$, even in the most extreme systems.
From this perspective, the HII~1132 and HD~22680 systems are notable, with 
excess ratios of 17.44 and 3.79 respectively reported by \citet{Sierchio+2010}. 
These values are much higher than any of the other dust belts in the sample 
and not expected in the passive, steady-state collisional evolution scenario at Pleiades age. 
The corresponding fractional luminosity for HII~1132, the largest outlier, is 
$1.9\times10^{-3}$ \citep[from Table 2 of][]{Sierchio+2010}. 
If the radius of this belt is 5~au, then it falls below the ALMA detection threshold at 
the sensitivity of our 1.3~mm survey (although it would have been readily detected at 70$~\mu$m
were it included in the FEPS sample for observation at that wavelength).
These systems are candidates of interest for deeper observations at longer wavelengths, 
to characterize their excess emission and to understand their physical nature. 
One intriguing possibility is that dynamical disturbances in the associated planetary systems 
have resulted in an epoch of unusually high dust production. 

Given the absence of any ALMA detections, consistency of the observations with the population 
synthesis prediction is not especially constraining to the collisional evolution model. 
An implicit assumption that remains to be explored is whether or not progenitors of debris disks 
in star clusters like the Pleiades could be different than those for the field population
near the Sun. The recent recognition of a high incidence of bright debris disks in F-type stars 
in the nearby 23~Myr-old $\beta$~Pictoris moving group compared with older moving groups and 
the field suggests a faster evolution than steady-state collisional evolution in the 
first $\sim100$~Myr, perhaps due to different initial disk properties or perhaps other 
factors, such as the influence of planets \citep{Pawellek+2021}. 
Since most field stars are thought to form in groups with $<100$ members like the 
nearby moving groups \citep{AdamsMyers2001}, it seems unlikely that initial conditions 
are responsible. By contrast, about 10\% of stars form in dense open clusters like the 
Pleiades, where it is more plausible that birth environment could be important. 
Studies of protoplanetary disk properties in rich cluster environments, in particular
the Orion Nebula Cluster, hint at different disk bulk properties than in more distributed 
star-forming regions like Taurus, in particular smaller sizes and lower masses in close
proximity to the harsh radiation fields of O-type stars \citep[e.g.,][]{Eisner+2018}.
If birth environment is not responsible, then the consistency of the Pleiades results 
with the population synthesis models suggests that whatever process results in overbright 
disks at 23~Myr has ended by $\sim$100~Myr.

\section{Conclusions}
\label{sec:conclusions}
The 115~Myr-old Pleiades cluster provides an important testbed for models of debris disk evolution.
Of particular interest are Solar-type stars surrounded by possible analogs of the early Kuiper Belt.
We used ALMA to survey 76 FGK type stars in the Pleiades at 1.3~mm to improve on previous observations 
at this long wavelength in both sensitivity and number to probe the presence of massive cold dust belts.
These observations obtained a typical beam size $\sim1\farcs5$ (200~au) and a median rms noise 54$~\mu$Jy,
and they showed no significant detections of circumstellar dust.
Several sources detected within the ALMA fields of view of the Pleiades stars are consistent 
with the expected background of dusty galaxies. 
For a dust belt at 40~K, the $3\sigma$ limit on 1.3~mm flux density corresponds to a 
fractional luminosity, $L_{dust}/L_{star} <10^{-4}$. 
These sensitive 1.3~mm observations provide a testbed for debris disk evolutionary models 
for a population at an age where there are few existing constraints.
A standard, passive collisional cascade model 
for steady state evolution is statistically consistent with the ALMA 1.3~mm flux density limits, 
as well as previous detection statistics at 24~$\mu$m and shorter infrared wavelengths. 
The millimeter observations reveal no outliers from the model expectations, in particular
no high fractional luminosity cold belts similar to those found around a few Solar-type stars 
in the local field population. Two of the Pleiades systems shows extreme $24~\mu$m excess 
emission that suggest significant departures from steady state evolution, perhaps resulting
from dynamical disturbances involving planets. 

\begin{acknowledgements}
This paper makes use of the following ALMA data: ADS/JAO.ALMA\#2019.1.00251.S. 
ALMA is a partnership of ESO (representing its member states), NSF (USA) and NINS (Japan),  together with 
NRC (Canada), MOST and ASIAA (Taiwan), and KASI (Republic of Korea), in cooperation with the Republic of Chile. 
The Joint ALMA Observatory is operated by ESO, AUI/NRAO and NAOJ.
The National Radio Astronomy Observatory is a facility of the National Science Foundation operated under 
cooperative agreement by Associated Universities, Inc.
D.S. thanks the Smithsonian Astrophysical Observatory for summer research support to carry out parts of this project. 
L.M. acknowledges funding from the European Union’s Horizon 2020 research and innovation programme under the Marie Sklodowska-Curie grant agreement No. 101031685.
MAM acknowledges the National Aeronautics and Space Administration 
under award number 19-ICAR19\_2-0041.
We thank Scott Kenyon and Qizhou Zhang for many valuable comments on an early version of the manuscript.
We also thank Joan Najita for stimulating discussion.
\end{acknowledgements}

\software{CASA (McMullin et al. 2007), 
UVMultiFit (Marti-Vidal et al. 2014),
astropy (The Astropy Collaboration 2013, 2018)}

\bibliography{pleiades}{}
\bibliographystyle{aasjournal}

\end{document}